\documentclass[letterpaper,11pt]{article}
\usepackage{amsmath,mathrsfs,amssymb,amsfonts,amsthm,mathtools}
\usepackage{physics,color}
\usepackage{graphicx}
\usepackage{subfigure}
\usepackage{tabularx} 
\usepackage[table]{xcolor}
\usepackage{tikz,tikz-cd}
\usetikzlibrary{shapes.geometric, arrows, positioning}
\usepackage{empheq}
\usepackage{bm}

\pdfoutput=1 
\usepackage{jheppub} 
\usepackage[T1]{fontenc} 

\newcommand{\nn}{\nonumber}

\def\ft#1#2{{\textstyle{\frac{\scriptstyle #1}{\scriptstyle #2} } }}
\def\fft#1#2{{\frac{#1}{#2}}}

\def\im{{{\rm i\,}}}

\hypersetup{
    colorlinks=true,
    linkcolor=blue,
    citecolor=blue,
    urlcolor=blue
}

\title{Kappa Plane Wave Modes and Continuous Squeezing in Quantum Field Theory}

\author[a]{Arash Azizi}

\affiliation[a]{{\it The Institute for Quantum Science and Engineering,
Texas A\&M University,\\ College Station, TX 77843, U.S.A.}}

\emailAdd{sazizi@tamu.edu}

\abstract{
We introduce a new family of field modes in flat spacetime—termed \textit{$\kappa$-plane wave modes}—constructed from $\kappa$-dependent linear combinations of Minkowski plane waves. These modes define a one-parameter family of vacua $\ket{0_\kappa}$ that smoothly interpolate between different quantizations, reducing to the Minkowski vacuum in the limit $\kappa \to 0$. We show that $\ket{0_\kappa}$ is uniquely characterized as a continuous-mode squeezed vacuum, with frequency-dependent squeezing parameter $r(\nu)$ satisfying $\tanh r(\nu) = e^{-\pi \nu/\kappa}$. We also derive two Bogoliubov transformations between $\kappa$-plane wave and $\kappa$-Rindler operators, which exhibit a universal form and smoothly interpolate between all known mode decompositions, including those of Minkowski, Rindler, and Unruh quantizations as limiting cases.
}

\begin{document} 
\maketitle
\flushbottom

\section{Introduction}

Quantum Field Theory (QFT) in curved spacetime \cite{DEWITT1975,Wald:1975kc, Fulling:1989nb, BirrellDavies1982, Wald1994, Mukhanov2007, Witten22QFT} has profoundly reshaped our understanding of quantum phenomena in non-inertial frames and gravitational backgrounds. Seminal results such as Hawking radiation \cite{Hawking1975} and the Unruh effect \cite{Unruh1976} highlight the essential observer-dependence of particle content and vacuum structure. These insights extend beyond high-energy and gravitational physics, influencing relativistic quantum information \cite{Landulfo2009, MannRalph2012, Aspling:2024wbz}, quantum optics \cite{Scully2003, Scully2006, Scully2011lasercooling}, and analog gravity systems \cite{Barcelo2011, Nation2012Nori}.

A cornerstone realization in this field is the absence of a unique, observer-independent vacuum state. Fulling’s pioneering work \cite{Fulling1973} established that inertial and non-inertial observers may assign different particle content to the same quantum field. This led to Unruh’s construction of mode solutions—now called Unruh modes—tailored to uniformly accelerated observers \cite{Unruh1976, UnruhWald1984}, with the celebrated prediction that such observers perceive the Minkowski vacuum as a thermal state at temperature
\begin{align}
T_U = \frac{\hbar a}{2\pi c k_B} ,.
\end{align}

In previous work \cite{Azizi2022kappashort, Azizi2023}, we introduced a novel class of mode functions—then dubbed \textit{$\kappa$-modes}—which combine Rindler modes from opposing wedges with $\kappa$-dependent weights. These modes gave rise to a family of vacua $\ket{0_\kappa}$ that interpolate between the Unruh vacuum ($\kappa=1$) and the Rindler vacuum ($\kappa \to 0$). To avoid ambiguity in the present work, we now refer to these as \textit{$\kappa$-Rindler modes}, with associated annihilation operators ${\cal B}_{\Omega,\kappa}$.

Here, we construct an entirely distinct set of $\kappa$-dependent modes from Minkowski plane waves—linear combinations of positive and negative frequency modes with $\kappa$-weighted coefficients. We refer to these as \textit{$\kappa$-plane wave modes}, with corresponding operators ${\cal A}{\Lambda,\kappa}$. The associated vacua, $\ket{0\kappa}$, approach the Minkowski vacuum as $\kappa \to 0$ but do not reduce to the Rindler vacuum in any limit. This construction thus yields a complementary family of quantum field representations in flat spacetime.

Crucially, we demonstrate that the $\kappa$-plane wave vacuum is a continuous-mode squeezed vacuum. Each frequency mode $\nu$ satisfies the equation
\begin{align}
\left( a_\nu - e^{-\frac{\pi \nu}{\kappa}} a^\dagger_\nu \right) \ket{0_\kappa} = 0,
\end{align}
indicating that $\ket{0_\kappa}$ is a pure Gaussian state with a frequency-dependent squeezing parameter. This places the entire construction within the formalism of continuous-variable (CV) quantum field theory and opens the door to applications of Gaussian quantum information techniques.

Another core result of this paper is the derivation of a pair of Bogoliubov transformations—between $\kappa$-plane wave and $\kappa$-Rindler modes—that generalize and unify all standard mode transformations in flat spacetime. These transformations interpolate continuously between Minkowski, Unruh, and Rindler quantizations and are structurally symmetric under $\kappa \leftrightarrow \kappa'$. Due to this universality and interpolating power, we refer to them as the \textit{Mother of All Bogoliubov Transformations}.

For clarity, we summarize the defining features of these modes and their associated vacua in the following table:


\begin{table}[h!]
\centering
\caption{Summary of kappa modes and their special limits.}
\label{tab:kappa_modes}
\arrayrulecolor{black}
\setlength{\arrayrulewidth}{0.3mm}
\renewcommand{\arraystretch}{1.5}
\begin{tabular}{|>{\columncolor{gray!15}}m{3.2cm}|
                 >{\centering\arraybackslash}m{2.5cm}|
                 m{4.1cm}| 
                 m{3.7cm}|} 
\hline
\rowcolor{blue!18}
\textbf{Mode} & \textbf{Operator} & \textbf{Special limit modes} & \textbf{Special limit vacua} \\
\hline
\textit{Kappa plane wave} 
& 
${\cal A}_{\Lambda,\kappa}$ \newline $(\Lambda \in \mathbb{R}^+)$ 
& 
\begin{tabular}[t]{r @{:\enspace} l}
$\kappa \to 0$ & Plane wave
\end{tabular}
& 
\begin{tabular}[t]{r @{:\enspace} l}
$\kappa \to 0$ & Minkowski
\end{tabular}
\\
\hline
\textit{Kappa Rindler}
& 
${\cal B}_{\Omega,\kappa}$ \newline $(\Omega \in \mathbb{R})$
& 
\begin{tabular}[t]{r @{:\enspace} l}
$\kappa \to 0$ & Rindler mode \\
$\kappa =1$ & Unruh mode
\end{tabular}
& 
\begin{tabular}[t]{r @{:\enspace} l}
$\kappa \to 0$ & Rindler \\
$\kappa=1$ & Minkowski
\end{tabular}
\\
\hline
\end{tabular}
\end{table} 

Table~\ref{tab:kappa_modes} summarizes the key features of the $\kappa$-plane wave and $\kappa$-Rindler modes, including their operators and the special limits they reduce to, as derived in Sections~\ref{kappa.planewave} and \ref{Bogol}.

The paper is organized as follows. Section~\ref{kappa.planewave} introduces the \(\kappa\)-plane wave modes. Section~\ref{Bogol} derives the Bogoliubov transformations connecting these modes to other known bases, culminating in a universal transformation we call the \textit{Mother of All Bogoliubov Transformations}. Section~\ref{com.rel.kappa.planewave} explores the deformed commutation relations, while Section~\ref{sec.cont.squeezing} presents our central result: the characterization of the \(\kappa\)-vacuum as a continuous-mode squeezed state. We conclude in Section~\ref{conclusion}. Four appendices provide technical support, justifying our ansatz and detailing key calculations.

\section{Kappa plane wave: combining plane wave modes with opposite sign norms} \label{kappa.planewave}

This work considers a massless scalar field propagating in (1+1)-dimensional Minkowski spacetime. We adopt natural units ($c=1$) and use light-cone coordinates defined by $u=t-x$ and $v=t+x$.
\subsection{Combination of plane wave Minkowski}

Consider the field mode defined as a combination of plane wave Minkowski modes as follows
\begin{equation}
\Phi(u,\Omega)=\alpha(\Omega) e^{-\im \Omega u}+\beta(\Omega) e^{\im \Omega u}\,,\label{mode}
\end{equation}
where $\Omega \in \mathbb{R}$. The inner product should satisfy the following
\begin{equation}
\left\langle \Phi(x, \Omega),\Phi(x, \Omega')\right\rangle=[a_\Omega,a^\dagger_{\Omega'}]=\delta(\Omega-\Omega')\,, \quad \quad
\left\langle \Phi(u, \Omega),\Phi^*(u, \Omega')\right\rangle=-\left[a_\Omega, a_ {\Omega^{\prime}}\right]=0\,. \label{innerprodrelations}
\end{equation}
Substituting the modes from (\ref{mode}), we have
\begin{align}
\left\langle\Phi(u, \Omega), \Phi\left(u, \Omega^{\prime}\right)\right\rangle =&\left\langle\alpha(\Omega) e^{-\im \Omega u}+\beta(\Omega) e^{\im \Omega u}, \alpha(\Omega') e^{-\im \Omega' u}+\beta(\Omega') e^{\im \Omega' u} \right\rangle \nn\\
=&4 \pi \Omega \delta\left(\Omega-\Omega^{\prime}\right)\left(\alpha^*(\Omega) \alpha\left(\Omega^{\prime}\right)
-\beta^*(\Omega) \beta\left(\Omega^{\prime}\right)\right) \nn\\
&+ 4 \pi \Omega \delta\left(\Omega+\Omega^{\prime}\right)\left(
\alpha^*(\Omega) \beta\left(\Omega^{\prime}\right)-\beta^*(\Omega) \alpha\left(\Omega^{\prime}\right)\right) \nn\\
=& 4 \pi \Omega \delta\left(\Omega-\Omega^{\prime}\right)\left(\abs{\alpha(\Omega)}^2
-\abs{\beta(\Omega)}^2\right)  \nn\\
&+ 4 \pi \Omega \delta\left(\Omega+\Omega^{\prime}\right)
\Big(\alpha^*(\Omega) \beta(-\Omega)
-\beta^*(\Omega) \alpha(-\Omega)\Big)\,, \label{inn.prod1}
\end{align}
where we have used the following result from the inner product
\begin{equation}
\left\langle 
e^{-\im \Omega u}, e^{-\im \Omega' u}
\right\rangle=4\pi \Omega\, \delta(\Omega-\Omega')\,. \label{inner-product}
\end{equation}
Therefore, demanding $\left\langle \Phi(x, \Omega),\Phi(x, \Omega')\right\rangle=\delta(\Omega-\Omega')$ yields
\begin{equation}
|\alpha(\Omega)|^2 -|\beta(\Omega)|^2=\frac{1}{4 \pi \Omega}\,, \label{constraint1}
\end{equation}
and 
\begin{equation}
\alpha^*(\Omega) \beta(-\Omega)
-\beta^*(\Omega) \alpha(-\Omega)=0\,,\label{constraint2}
\end{equation}
if $\delta\left(\Omega+\Omega^{\prime}\right)$ were to be nonzero. 

Next, we consider the inner product between a mode and the complex conjugate of a mode.  Requiring $\left\langle \Phi^*(x, \Omega),\Phi(x, \Omega')\right\rangle=0$ results
\begin{align}  
\left\langle\Phi^*(u, \Omega), \Phi\left(u, \Omega^{\prime}\right)\right\rangle =&\left\langle\alpha^*(\Omega) e^{\im \Omega u}+\beta^*(\Omega) e^{-\im \Omega u}, 
\alpha(\Omega') e^{-\im \Omega' u}+\beta(\Omega') e^{\im \Omega' u} \right\rangle \nn\\
=&4 \pi \Omega \delta\left(\Omega-\Omega^{\prime}\right)
\left(\beta(\Omega) \alpha\left(\Omega^{\prime}\right)-\alpha(\Omega) \beta\left(\Omega^{\prime}\right)
\right) \nn\\
&+ 4 \pi \Omega \delta\left(\Omega+\Omega^{\prime}\right) \left(
-\alpha(\Omega) \alpha\left(\Omega^{\prime}\right)
+\beta(\Omega) \beta\left(\Omega^{\prime}\right)\right) \nn\\
=& 4 \pi \Omega \delta\left(\Omega+\Omega^{\prime}\right)\Big(
-\alpha(\Omega) \alpha\left(-\Omega\right)
+\beta(\Omega) \beta\left(-\Omega\right)\Big)\,, \label{inn.prod2}
\end{align}
which indicates 
\begin{equation}
-\alpha(\Omega) \alpha\left(-\Omega\right)
+\beta(\Omega) \beta\left(-\Omega\right)=0\,, \label{constraint3}
\end{equation}
if $\delta\left(\Omega+\Omega^{\prime}\right)$ were to be nonzero.

Thus, the inner product structure imposes three constraints. By combining (\ref{constraint2}) and (\ref{constraint3}), we find
\begin{equation}
\alpha\left(-\Omega\right)\beta\left(-\Omega\right)\Big(|\alpha(\Omega)|^2-|\beta(\Omega)|^2\Big)=0\,. \label{constraint4}
\end{equation}
However, since the normalization constraint (\ref{constraint1}) forbids $|\alpha(\Omega)|^2 - |\beta(\Omega)|^2 = 0$, the only way to satisfy (\ref{constraint4}) is to demand $\alpha(-\Omega)\beta(-\Omega)=0$. This leads to a contradiction: if $\alpha(\Omega)$ vanishes, so must $\beta(\Omega)$, and vice versa, which is inconsistent with the normalization condition.

This contradiction suggests that allowing $\Omega$ to span the full real line $\mathbb{R}$ leads to inconsistency. A natural resolution is to restrict to $\Omega > 0$. In this regime, the condition $\delta\left(\Omega+\Omega^{\prime}\right)=0$ is automatically satisfied, eliminating the need to impose (\ref{constraint2}) and (\ref{constraint3}).

Assuming $\Omega > 0$, we are left with only one constraint from (\ref{inn.prod1}), namely the normalization condition (\ref{constraint1}). Following \cite{Azizi2023}, we choose the following ansatz:
\begin{equation}
\alpha(\Lambda)=\frac{e^{\fft{\pi \Lambda}{2\kappa}}}{\sqrt{8 \pi \Lambda\, \sinh{(\fft{\pi \Lambda}{\kappa} )}}} \,,
\qquad \qquad
\beta(\Lambda)=\frac{e^{\fft{-\pi \Lambda}{2\kappa}}}{\sqrt{8 \pi \Lambda\, \sinh{(\fft{\pi \Lambda}{\kappa} )}}}\,. \label{ansatz}
\end{equation}
While the normalization condition (\ref{constraint1}) allows for an infinite family of solutions, the ansatz presented in Eq. (\ref{ansatz}) is not arbitrary. This specific form is required for two crucial physical reasons. First, it ensures the parameter $\kappa$ is a frequency-independent constant. Second, and most importantly, this is the unique choice that allows the resulting $\kappa$-vacuum to be interpreted as a continuous-mode squeezed state of the Minkowski vacuum. This physical insight is the central motivation for our construction. A more detailed technical justification is provided in the Appendix \ref{app.ansatz}.

With the above mentioned ansatz, the mode becomes 
\vspace{.35cm}
\begin{equation}
\boxed{\quad \Phi(u,\Lambda,\kappa)= \fft1{{\sqrt{8 \pi \Lambda\, \sinh{(\fft{\pi \Lambda}{\kappa} )}}}}
\left\{e^{\fft{\pi \Lambda}{2\kappa}}\,
e^{-\im \Lambda u}
+e^{-\fft{\pi \Lambda}{2\kappa}}\,e^{\im \Lambda u} \right\} \,. \quad }\label{kappamodeplanewave}
\end{equation}
This mode, referred to as the \textit{kappa plane wave mode}, combines positive and negative frequency Minkowski plane waves with $\kappa$-dependent weights, enabling a continuous interpolation to the standard Minkowski mode in the limit $\kappa \to 0$, as shown in Eq.~\eqref{PlanewaveMink}. The weights $e^{\fft{\pi \Lambda}{2\kappa}}$ and $e^{-\fft{\pi \Lambda}{2\kappa}}$ introduce a frequency-dependent squeezing effect, which we demonstrate in Section~\ref{sec.cont.squeezing} leads to the vacuum $\ket{0_\kappa}$ being a continuous-mode squeezed state relative to the Minkowski vacuum. The field then can be written as
\begin{equation}
\Phi(u)=\int_{0}^{\infty} d\Lambda \Big(\Phi(u,\Lambda,\kappa) {\cal A}_{\Lambda,\kappa} + \Phi^*(u,\Lambda,\kappa) {\cal A}^{\dagger}_{\Lambda,\kappa} \Big)\,,
\end{equation}
where ${\cal A}_{\Lambda,\kappa}$ represents the annihilation operator for a $\kappa$-mode with the frequency $\Lambda$.  Explicitly, using (\ref{kappamodeplanewave}) one has
\begin{align} 
\Phi(u) = \int_{0}^{\infty} d\Lambda 
 \fft1{{\sqrt{8 \pi \Lambda\, \sinh{(\fft{\pi \Lambda}{\kappa} )}}}} 
\bigg\{&\Big( e^{\fft{\pi \Lambda}{2\kappa}}\,
e^{-\im \Lambda u}
+e^{-\fft{\pi \Lambda}{2\kappa}}\,e^{\im \Lambda u} \Big)\,
{\cal A}_{\Lambda,\kappa} \nn\\
&+\Big(e^{\fft{\pi \Lambda}{2\kappa}}\,
e^{\im \Lambda u}
+e^{-\fft{\pi \Lambda}{2\kappa}}\,e^{-\im \Lambda u} \Big)\,
{\cal A}^{\dagger}_{\Lambda,\kappa}\bigg\} \,. \label{kappaplanewavefield} 
\end{align}

\subsection{Minkowski plane wave as a special case}

Consider the limit $\kappa \rightarrow 0$ in the expression (\ref{kappamodeplanewave}). Since both $\Lambda$ and $\kappa$ are strictly positive, the exponential suppression in the second term causes it to vanish, and the mode reduces to
\begin{equation}
\Phi(u, \Lambda > 0,\kappa\rightarrow 0)= \fft1{\sqrt{4 \pi \Lambda}}
e^{-\im \Lambda u} \,. \label{PlanewaveMink}
\end{equation}
This corresponds to the standard positive-frequency Minkowski plane wave. Hence, in the limit $\kappa \to 0$, the $\kappa$-mode smoothly recovers the Minkowski mode structure. Moreover, we have
\begin{equation}
{\cal A}_{\Lambda,0}=a_{\Lambda} \,. 
\end{equation}

\section{Bogoliubov transformations} \label{Bogol}

\subsection{Bogoliubov transformation between kappa plane wave modes}
We now derive the Bogoliubov transformation between two distinct $\kappa$-plane wave mode sets. Let the field $\Phi(u)$ be expanded in both $\kappa$ and $\kappa'$ bases as follows:
\begin{align} 
\Phi(u) =& \int_{0}^{\infty} d\Lambda 
 \fft1{{\sqrt{8 \pi \Lambda\, \sinh{(\fft{\pi \Lambda}{\kappa} )}}}} 
\Big\{\Big( e^{\fft{\pi \Lambda}{2\kappa}}\,
e^{-\im \Lambda u}
+e^{-\fft{\pi \Lambda}{2\kappa}}\,e^{\im \Lambda u} \Big)\,
{\cal A}_{\Lambda,\kappa} \nn\\
& \phantom{\int_{0}^{\infty} d\Lambda 
 \fft1{{\sqrt{8 \pi \Lambda\, \sinh{(\fft{\pi \Lambda}{\kappa} )}}}} 
\Big\{}
+\Big(e^{\fft{\pi \Lambda}{2\kappa}}\,
e^{\im \Lambda u}
+e^{-\fft{\pi \Lambda}{2\kappa}}\,e^{-\im \Lambda u} \Big)\,
{\cal A}^{\dagger}_{\Lambda,\kappa}\Big\} \nn\\
=& \int_{0}^{\infty} d\Lambda 
 \fft1{{\sqrt{8 \pi \Lambda\, \sinh{(\fft{\pi \Lambda}{\kappa'} )}}}} 
\Big\{\Big( e^{\fft{\pi \Lambda}{2\kappa'}}\,
e^{-\im \Lambda u}
+e^{-\fft{\pi \Lambda}{2\kappa'}}\,e^{\im \Lambda u} \Big)\,
{\cal A}_{\Lambda,\kappa'} \nn\\
&\phantom{\int_{0}^{\infty} d\Lambda 
 \fft1{{\sqrt{8 \pi \Lambda\, \sinh{(\fft{\pi \Lambda}{\kappa'} )}}}} 
\Big\{}
+\Big(e^{\fft{\pi \Lambda}{2\kappa'}}\,
e^{\im \Lambda u}
+e^{-\fft{\pi \Lambda}{2\kappa'}}\,e^{-\im \Lambda u} \Big)\,
{\cal A}^{\dagger}_{\Lambda,\kappa'}\Big\}
\,. \label{bogol1} 
\end{align}
To obtain the transformation, we equate the coefficients of $e^{-\im \Lambda u}$ and $e^{\im \Lambda u}$ on both sides of (\ref{bogol1}). Matching the coefficients of $e^{-\im \Lambda u}$ yields
\begin{equation}
\fft1{{\sqrt{8 \pi \Lambda\, \sinh{(\fft{\pi \Lambda}{\kappa} )}}}} 
\left\{e^{\fft{\pi \Lambda}{2\kappa}} {\cal A}_{\Lambda,\kappa}
+e^{-\fft{\pi \Lambda}{2\kappa}} {\cal A}^{\dagger}_{\Lambda,\kappa}\right\}
=\fft1{{\sqrt{8 \pi \Lambda\, \sinh{(\fft{\pi \Lambda}{\kappa'} )}}}} 
\left\{e^{\fft{\pi \Lambda}{2\kappa'}}
{\cal A}_{\Lambda,\kappa'}
+e^{-\fft{\pi \Lambda}{2\kappa'}}\,
{\cal A}^{\dagger}_{\Lambda,\kappa'}\right\}\,, \label{bogol2}
\end{equation}
while comparing the coefficients of $e^{\im \Lambda u}$ gives
\begin{equation}
\fft1{{\sqrt{8 \pi \Lambda\, \sinh{(\fft{\pi \Lambda}{\kappa} )}}}} 
\left\{e^{-\fft{\pi \Lambda}{2\kappa}} {\cal A}_{\Lambda,\kappa}
+e^{\fft{\pi \Lambda}{2\kappa}} {\cal A}^{\dagger}_{\Lambda,\kappa}\right\}
=\fft1{{\sqrt{8 \pi \Lambda\, \sinh{(\fft{\pi \Lambda}{\kappa'} )}}}} 
\left\{e^{-\fft{\pi \Lambda}{2\kappa'}}
{\cal A}_{\Lambda,\kappa'}
+e^{\fft{\pi \Lambda}{2\kappa'}}\,
{\cal A}^{\dagger}_{\Lambda,\kappa'}\right\}\,.\label{bogol3}
\end{equation}

Solving (\ref{bogol2}) and (\ref{bogol3}) simultaneously, we find the Bogoliubov transformation between operators corresponding to different $\kappa$-modes:
\begin{equation}
{\cal A}_{\Lambda,\kappa'}=
\fft{1}{\sqrt{{\sinh{(\fft{\pi \Lambda}{\kappa} )}\sinh{(\fft{\pi \Lambda}{\kappa'} )}}}}
\Bigg\{ \sinh{\Big(\ft{\pi \Lambda}2\,(\ft1{\kappa'}+\ft1{\kappa} \big)}\Big) \,{\cal A}_{\Lambda,\kappa}
+ \sinh{\Big(\ft{\pi \Lambda}2\,(\ft1{\kappa'}-\ft1{\kappa} \big)}\Big) \,{\cal A}^{\dagger}_{\Lambda,\kappa} 
\Bigg\}\,. \label{bogolfinal}
\end{equation}

This expression shows that the annihilation operator ${\cal A}_{\Lambda,\kappa'}$ includes both annihilation and creation operators from the $\kappa$-basis, indicating that the two sets of modes are related through a nontrivial Bogoliubov transformation. Consequently, the vacua associated with different values of $\kappa$ are inequivalent.

\subsubsection{Bogoliubov transformation between a  kappa plane wave  and the Minkowski plane wave} \label{Bogol.kappa-plane.Mink}
It is straightforward to derive the Bogoliubov transformation between a $\kappa$-plane wave mode and the standard Minkowski plane wave. This can be achieved by taking the limit $\kappa \rightarrow 0$ in equation (\ref{bogolfinal}), which yields
\begin{equation}
{\cal A}_{\Lambda,\kappa}=
\fft{1}{\sqrt{{2\sinh{(\fft{\pi \Lambda}{\kappa} )}}}}
\Big( e^{\fft{\pi \Lambda}{2\kappa}} \,a_{\Lambda}
- e^{\fft{-\pi \Lambda}{2\kappa}} \,a^{\dagger}_{\Lambda} 
\Big)\,. \label{Kappa planewave-Mink.}
\end{equation}
\subsection{Kappa plane wave and Rindler} \label{Bogol.kappa-plane.Rind}

In this section, we explore the relationship between the $\kappa$-plane wave modes and the Rindler modes by computing the Bogoliubov transformation connecting them. We begin with the $\kappa$-plane wave mode expansion of the field:
\begin{align}
\Phi(u) &=\int_0^{\infty} d \Omega
\frac{1}{\sqrt{8 \pi \Omega \sinh \left(\frac{\pi \Omega}{k}\right)}}  \Bigg\{
\left(e^{-\frac{\pi \Omega}{2 \pi}} e^{\im \Omega u}+e^{\frac{\pi \Omega}{2 k}} e^{-\im \Omega u}\right) {\cal A}_{\Omega,\kappa} \nn\\
&\phantom{=\int_0^{\infty} d \Omega
\frac{1}{\sqrt{8 \pi \Omega \sinh \left(\frac{\pi \Omega}{k}\right)}}  \Bigg\{}
+\left(e^{-\frac{\pi \Omega}{2 k}} e^{-\im \Omega u}+e^{\frac{\pi \Omega}{2 k}} e^{\im \Omega u}\right) {\cal A}_{\Omega,\kappa}^{\dagger}\Bigg\}   \nn\\
&=\theta(-u)\,\int_{0}^{\infty}
\fft{d \Omega}{\sqrt{4 \pi \Omega}}\,
\Big ( (-au)^{\ft{\im \Omega}a}\,b_{R\, \Omega}
+(-au)^{-\ft{\im \Omega}a}\,b_{R\, \Omega}^\dagger \Big) \nn\\
&\phantom{=}
+\theta(u)\,\int_{0}^{\infty}
\fft{d \Omega}{\sqrt{4 \pi \Omega}}\,
\Big ( (au)^{\ft{-\im \Omega}a}\,b_{L\, \Omega}
+(au)^{\ft{\im \Omega}a}\,b_{L\, \Omega}^\dagger \Big)\,.   \label{RindKappaPW-Bogol1}
\end{align}
Now finding out $\int_{-\infty}^{\infty} d u\, e^{\im \Lambda u} \Phi(u)$, where $\Lambda>0$ yields
\begin{align}
\int_{-\infty}^{\infty} d u\, e^{\im \Lambda u} \Phi(u) =&
\frac{2 \pi}{\sqrt{8 \pi \Lambda \sinh \left(\frac{\pi 
\Lambda}{\kappa}\right)}}\left(e^{\frac{\pi \Lambda}{2 \kappa}} {\cal A}_{\Lambda,\kappa}
+e^{-\frac{\pi \Lambda}{2 \kappa}} {\cal A}^{\dagger}_{\Lambda,\kappa}\right)  \nn\\
=&\im \int_{0}^{\infty} \fft{d \Omega}{\sqrt{4 \pi \Omega}}\,
\Bigg \{-a^{\frac{\im \omega}{a}} \Lambda^{-\left(1+\frac{\im \omega}{a}\right)}\, e^{\frac{\pi \omega}{2 a}}\, \Gamma \big(1+\ft{\im \omega}{a}\big)
\,b_{R\, \Omega} \nn\\
&\phantom{\im \int_{0}^{\infty} \fft{d \Omega}{\sqrt{4 \pi \Omega}}\,
\Bigg \{}
-a^{-\frac{\im \omega}{a}} \Lambda^{-\left(1-\frac{\im \omega}{a}\right)}\, e^{-\frac{\pi \omega}{2 a}}\, \Gamma \big(1-\ft{\im \omega}{a}\big)
\,b^{\dagger}_{R\, \Omega} \nn\\
&\phantom{\im \int_{0}^{\infty} \fft{d \Omega}{\sqrt{4 \pi \Omega}}\,
\Bigg \{}
+a^{-\frac{\im \omega}{a}} \Lambda^{-\left(1-\frac{\im \omega}{a}\right)}\, e^{\frac{\pi \omega}{2 a}}\, \Gamma \big(1-\ft{\im \omega}{a}\big)
\,b_{L\, \Omega} \nn\\
&\phantom{\im \int_{0}^{\infty} \fft{d \Omega}{\sqrt{4 \pi \Omega}}\,
\Bigg \{}
+a^{\frac{\im \omega}{a}} \Lambda^{-\left(1+\frac{\im \omega}{a}\right)}\, e^{-\frac{\pi \omega}{2 a}}\, \Gamma \big(1+\ft{\im \omega}{a}\big)
\,b^{\dagger}_{L\, \Omega}
\Bigg\} \,,\label{RindKappaPW-Bogol2}
\end{align}
where the following useful integral has been utilized
\begin{align}
\int_{-\infty}^{+\infty} d u\, e^{\im \nu u}\, (-u)^{\im \Omega} \theta(-u) =&\int_{0}^{\infty} d u\, e^{-\im \nu u}\, u^{\im \Omega}=-\im \nu^{-(1+\im \Omega)} e^{\frac{\pi \Omega}{2}} \Gamma(1+\im \Omega) \,, \nn\\
\int_{-\infty}^{+\infty} d u\, e^{\im \nu u}\, u^{\im \Omega} \theta(u)=&\int_{0}^{\infty} d u\, e^{\im \nu u} u^{\im \Omega}=\im \nu^{-(1+\im \Omega)} e^{-\frac{\pi \Omega}{2}} \Gamma(1+\im \Omega) \,.
\end{align}
By performing the analogous calculation for $\Lambda<0$ and combining the two results, we isolate ${\cal A}_{\Lambda,\kappa}$ in terms of Rindler operators. This yields the final expression:
\begin{align}
{\cal A}_{\Lambda,\kappa}=&
\int_{0}^{+\infty} d \Omega\,\im
\fft{1}{\sqrt{2 \pi^2 \Lambda\Omega \sinh \left(\frac{\pi 
\Lambda}{\kappa}\right)}} \label{RindKappaPW-Bogol3}\\
& \times
\Bigg \{-a^{\frac{\im \omega}{a}} \Lambda^{-\frac{\im \omega}{a}}\, 
\Gamma \big(1+\ft{\im \omega}{a}\big)  \,
\Bigg( \cosh \left(\ft{\pi \Lambda}{2 \kappa}+\ft{\pi \omega}{2 a}\right)
b_{R\, \Omega} 
- \cosh \left(\ft{\pi \Lambda}{2 \kappa}-\ft{\pi \omega}{2 a}\right) 
b^{\dagger}_{L\, \Omega} \Bigg) \nn\\
&\phantom{ \times\Bigg \{..}
+a^{-\frac{\im \omega}{a}} \Lambda^{\frac{\im \omega}{a}}\,  \Gamma \big(1-\ft{\im \omega}{a}\big)
\Bigg(\cosh \left(\ft{\pi \Lambda}{2 \kappa}+\ft{\pi \omega}{2 a}\right)
b_{L\, \Omega} 
\,- \cosh \left(\ft{\pi \Lambda}{2 \kappa}-\ft{\pi \omega}{2 a}\right) b^{\dagger}_{R\, \Omega} 
\Bigg)\Bigg \}\,. \nn
\end{align}
Finally, taking the limit $\kappa \rightarrow 0$ in (\ref{RindKappaPW-Bogol3}), we recover the well-known Bogoliubov transformation between Minkowski and Rindler operators:
\begin{align} 
a_{\nu}=\int_{0}^{\infty} d \Omega\,
\fft{-\im }{2\pi \sqrt{\Omega \nu}}\,e^{\ft{\pi \Omega}{2a}}\,
\Bigg\{& \phantom{+.} (\ft a{\nu})^{\ft{\im \Omega}a} \quad \gamma(1+\ft{\im \Omega}a)\,
\Big (b_{R\, \Omega}- e^{\ft{-\pi \Omega}{a}}\,b_{L\, \Omega}^\dagger \Big) \nn\\
&-(\ft a{\nu})^{\ft{-\im \Omega}a}\, \gamma (1-\ft{\im \Omega}a)\,
\Big (b_{L\, \Omega}- e^{\ft{-\pi \Omega}{a}}\,b_{R\, \Omega}^\dagger \Big)\Bigg\}\,, \label{Mink-Rind-a}
\end{align}
which agrees with the standard result for Minkowski-Rindler mode decomposition.

\subsection{Kappa plane wave and Kappa Rindler} \label{Bogol.kappa-plane.kappa-Rind}

\subsubsection{Kappa plane wave in terms of Kappa Rindler} \label{KappaPWtoKappaRind}

To establish the Bogoliubov transformation between $\kappa$-plane wave modes and $\kappa$-Rindler modes, we begin by expressing the field $\Phi(u)$ in terms of both mode bases:
\begin{align} 
\Phi(u) =& \int_{0}^{\infty} d\Lambda 
 \fft1{{\sqrt{8 \pi \Lambda\, \sinh{(\fft{\pi \Lambda}{\kappa} )}}}} 
\Bigg\{ \phantom{+} \Big( e^{\fft{\pi \Lambda}{2\kappa}}\,
e^{-\im \Lambda u}
+e^{-\fft{\pi \Lambda}{2\kappa}}\,e^{\im \Lambda u} \Big)\,
{\cal A}_{\Omega,\kappa}  \label{KPW-KR1}\\
& \phantom{\int_{0}^{\infty} d\Omega 
 \fft1{{\sqrt{8 \pi \Lambda\, \sinh{(\fft{\pi \Lambda}{\kappa} )}}}} 
\Bigg\{}
+\Big(e^{\fft{\pi \Lambda}{2\kappa}}\,
e^{\im \Lambda u}
+e^{-\fft{\pi \Lambda}{2\kappa}}\,e^{-\im \Lambda u} \Big)\,
{\cal A}^{\dagger}_{\Omega,\kappa}\Bigg\} \nn\\
=& \, \theta(-u) \int_{-\infty}^{\infty} d\Omega 
 \fft1{{\sqrt{8 \pi \Omega\, \sinh{(\fft{\pi \Omega}{\kappa'} )}}}} 
\bigg\{(-u)^{\im \Omega}\, 
e^{\fft{\pi \Omega}{2\kappa'}}\, {\cal B}_{\Omega,\kappa'}
+(-u)^{-\im \Omega}\, 
e^{\fft{\pi \Omega}{2\kappa'}}\, 
{\cal B}^{\dagger}_{\Omega,\kappa'}\bigg\} \nn\\
&+\theta(u) \int_{-\infty}^{\infty} d\Omega 
 \fft1{{\sqrt{8 \pi \Omega\, \sinh{(\fft{\pi \Omega}{\kappa'} )}}}} 
\bigg\{ u^{\im \Omega}\, e^{-\fft{\pi \Omega}{2\kappa'}}\, 
{\cal B}_{\Omega,\kappa'} 
+ u^{-\im \Omega}\, e^{-\fft{\pi \Omega}{2\kappa'}}\, 
{\cal B}^{\dagger}_{\Omega,\kappa'}\bigg\}\,. 
\end{align}
Applying the Fourier transform $\int_{-\infty}^{+\infty} d u\, e^{\im \Sigma u} \Phi(u)$ with $\Sigma > 0$ to both representations yields:
\begin{align} 
\int_{-\infty}^{+\infty} d u e^{\im \Sigma u} \Phi(u) 
&=\frac{2 \pi}{\sqrt{8 \pi \Sigma \sinh \left(\frac{\pi \Sigma}{\kappa}\right)}}\left(e^{\frac{\pi \Sigma}{2 \kappa}}  {\cal A}_{\Sigma, \kappa}+e^{-\frac{\pi \Sigma}{2 \kappa}}  {\cal A}_{\Sigma, \kappa}^{\dagger}\right) \nn\\
&=\int_{-\infty}^{+\infty} d \Omega \frac{\im}{\sqrt{8 \pi \Omega \sinh \left(\frac{\pi \Omega}{\kappa^{\prime}}\right)}}
\Bigg\{
-\Sigma^{-(1+ \im \Omega)} e^{\frac{\pi \Omega}{2}} \Gamma(1+ \im \Omega) e^{\frac{\pi \Omega}{2 \kappa'}} {\cal B}_{\Omega, \kappa^{\prime}} \nn\\
&\phantom{=\int_{-\infty}^{+\infty} d \Omega \frac{\im}{\sqrt{8 \pi \Omega \sinh \left(\frac{\pi \Omega}{\kappa^{\prime}}\right)}}\Bigg\{}
-\Sigma^{-(1- \im \Omega)} e^{\frac{-\pi \Omega}{2}} \Gamma(1- \im \Omega) e^{\frac{\pi \Omega}{2 \kappa'}} {\cal B}^{\dagger}_{\Omega, \kappa^{\prime}} \nn\\
&\phantom{=\int_{-\infty}^{+\infty} d \Omega \frac{\im}{\sqrt{8 \pi \Omega \sinh \left(\frac{\pi \Omega}{\kappa^{\prime}}\right)}}\Bigg\{}
+\Sigma^{-(1+ \im \Omega)} e^{\frac{-\pi \Omega}{2}} \Gamma(1+ \im \Omega) e^{\frac{-\pi \Omega}{2 \kappa'}} {\cal B}_{\Omega, \kappa^{\prime}} \nn\\
&\phantom{=\int_{-\infty}^{+\infty} d \Omega \frac{\im}{\sqrt{8 \pi \Omega \sinh \left(\frac{\pi \Omega}{\kappa^{\prime}}\right)}}\Bigg\{}
+\Sigma^{-(1- \im \Omega)} e^{\frac{\pi \Omega}{2}} \Gamma(1- \im \Omega) e^{\frac{-\pi \Omega}{2 \kappa'}} {\cal B}^{\dagger}_{\Omega, \kappa^{\prime}} \Bigg\} \,,
\end{align}
By forming an appropriate linear combination of the above result and its Hermitian conjugate, we isolate the Bogoliubov coefficients and obtain:
\begin{align}
&{\cal A}_{\Sigma,\kappa}= \nn\\
&\int_0^{\infty} d \Omega \frac{\im}{2 \pi}   \frac{1}{\sqrt{\Sigma \Omega \sinh \left(\frac{\pi \Sigma}{\kappa}\right) \sinh \left(\frac{\pi \Omega}{\kappa^{\prime}}\right)}} \times \label{RindKappa-PWKappa-Bogol3}\\
&
\Bigg \{ -\Sigma^{-\im \Omega}\, 
\Gamma \big(1+\im \Omega\big) 
\Bigg(\phantom{+.}
\Big[ e^{\frac{\pi \Sigma}{2 \kappa}} \sinh \left(\ft{\pi \Omega}{2}(1+\frac{1}{\kappa^{\prime}})\right)
- e^{\frac{-\pi \Sigma}{2 \kappa}} \sinh \left(\ft{\pi \Omega}{2}(1-\frac{1}{\kappa^{\prime}})\right)
\Big] {\cal B}_{\Omega,\kappa'}  \nn\\
&\phantom{\Bigg \{ -\Sigma^{-\im \Omega}\, 
\Gamma \big(1+\im \Omega\big) 
\Bigg(} +
\Big[ e^{\frac{\pi \Sigma}{2 \kappa}} \sinh \left(\ft{\pi \Omega}{2}(1-\frac{1}{\kappa^{\prime}})\right)
- e^{\frac{-\pi \Sigma}{2 \kappa}} \sinh \left(\ft{\pi \Omega}{2}(1+\frac{1}{\kappa^{\prime}})\right)
\Big] {\cal B}^{\dagger}_{-\Omega,\kappa'} \Bigg)  \nn\\
&\phantom{\Bigg \{}
 +\Sigma^{\im \Omega}\, 
\Gamma \big(1-\im \Omega\big) \,\,
\Bigg( \phantom{+..} \Big[ e^{\frac{\pi \Sigma}{2 \kappa}} \sinh \left(\ft{\pi \Omega}{2}(1-\frac{1}{\kappa^{\prime}})\right)
- e^{\frac{-\pi \Sigma}{2 \kappa}} \sinh \left(\ft{\pi \Omega}{2}(1+\frac{1}{\kappa^{\prime}})\right)
\Big] {\cal B}^{\dagger}_{\Omega,\kappa'}  \nn\\
&\phantom{\Bigg \{ -\Sigma^{-\im \Omega}\, 
\Gamma \big(1+\im \Omega\big) 
\Bigg(} +
\Big[ e^{\frac{\pi \Sigma}{2 \kappa}} \sinh \left(\ft{\pi \Omega}{2}(1+\frac{1}{\kappa^{\prime}})\right)
- e^{\frac{-\pi \Sigma}{2 \kappa}} \sinh \left(\ft{\pi \Omega}{2}(1-\frac{1}{\kappa^{\prime}})\right)
\Big] {\cal B}_{-\Omega,\kappa'} \Bigg) \Bigg \}\,. \nn
\end{align}
By symmetrizing over the frequency domain, the final expression for the Bogoliubov transformation between $\kappa$-plane wave modes and $\kappa$-Rindler modes becomes
\setlength\fboxsep{15pt}
\begin{empheq}[box=\fbox]{align}
{\cal A}_{\Lambda,\kappa}=& 
 \frac{\im}{2 \pi}  \int_{-\infty}^{\infty}  d \Omega \,   \frac{1}{\sqrt{\Lambda \Omega \sinh \left(\frac{\pi \Lambda}{\kappa}\right) \sinh \left(\frac{\pi \Omega}{\kappa^{\prime}}\right)}} \label{KappaPlane.intermsof.KappaRindler}\\
&\times \Bigg \{ -\Lambda^{-\im \Omega}\, 
\Gamma \big(1+\im \Omega\big) 
\Bigg[
 e^{\frac{\pi \Lambda}{2 \kappa}} \sinh \left(\ft{\pi \Omega}{2}(1+\frac{1}{\kappa^{\prime}})\right)
- e^{-\frac{\pi \Lambda}{2 \kappa}} \sinh \left(\ft{\pi \Omega}{2}(1-\frac{1}{\kappa^{\prime}})\right)
\Bigg] {\cal B}_{\Omega,\kappa'}  \nn\\
&\phantom{\times \Bigg \{....} +
\Lambda^{\im \Omega}\, \Gamma \big(1-\im \Omega\big) 
\Bigg[
e^{\frac{\pi \Lambda}{2 \kappa}} \sinh \left(\ft{\pi \Omega}{2}(1-\frac{1}{\kappa^{\prime}})\right)
- e^{-\frac{\pi \Lambda}{2 \kappa}} \sinh \left(\ft{\pi \Omega}{2}(1+\frac{1}{\kappa^{\prime}})\right)
\Bigg] {\cal B}^{\dagger}_{\Omega,\kappa'} 
\Bigg \}\,. \nn
\end{empheq}

\subsubsection{Kappa Rindler in terms of Kappa plane wave} \label{KappaRindtoKappaPW}
We now derive the Bogoliubov transformation expressing $\kappa$-Rindler modes in terms of $\kappa$-plane wave modes. Starting again with the field $\Phi(u)$ written in both bases:
\begin{align} 
\Phi(u) =& \int_{0}^{\infty} d\Lambda 
 \fft1{{\sqrt{8 \pi \Lambda\, \sinh{(\fft{\pi \Lambda}{\kappa} )}}}} 
\Bigg\{ \phantom{+} \Big( e^{\fft{\pi \Lambda}{2\kappa}}\,
e^{-\im \Lambda u}
+e^{-\fft{\pi \Lambda}{2\kappa}}\,e^{\im \Lambda u} \Big)\,
{\cal A}_{\Omega,\kappa}  \nn\\
& \phantom{\int_{0}^{\infty} d\Omega 
 \fft1{{\sqrt{8 \pi \Lambda\, \sinh{(\fft{\pi \Lambda}{\kappa} )}}}} 
\Bigg\{}
+\Big(e^{\fft{\pi \Lambda}{2\kappa}}\,
e^{\im \Lambda u}
+e^{-\fft{\pi \Lambda}{2\kappa}}\,e^{-\im \Lambda u} \Big)\,
{\cal A}^{\dagger}_{\Omega,\kappa}\Bigg\} \nn\\
=& \, \theta(-u) \int_{-\infty}^{\infty} d\Omega 
 \fft1{{\sqrt{8 \pi \Omega\, \sinh{(\fft{\pi \Omega}{\kappa'} )}}}} 
\left\{(-u)^{\im \Omega}\, 
e^{\fft{\pi \Omega}{2\kappa'}}\, {\cal B}_{\Omega,\kappa'}
+(-u)^{-\im \Omega}\, 
e^{\fft{\pi \Omega}{2\kappa'}}\, 
{\cal B}^{\dagger}_{\Omega,\kappa'}\right\} \nn\\
&+\theta(u) \int_{-\infty}^{\infty} d\Omega 
 \fft1{{\sqrt{8 \pi \Omega\, \sinh{(\fft{\pi \Omega}{\kappa'} )}}}} 
\left\{ u^{\im \Omega}\, e^{-\fft{\pi \Omega}{2\kappa'}}\, 
{\cal B}_{\Omega,\kappa'} 
+ u^{-\im \Omega}\, e^{-\fft{\pi \Omega}{2\kappa'}}\, 
{\cal B}^{\dagger}_{\Omega,\kappa'}\right\}\,. \label{KR-KPW1}
\end{align}
To isolate the $\kappa$-Rindler operators, we apply the weighted integrals 
\[
\int_{-\infty}^{+\infty} d u \, \theta(u)\, u^{\im \Sigma - 1} \Phi(u)
\quad \text{and} \quad 
\int_{-\infty}^{+\infty} d u \, \theta(-u)\, (-u)^{\im \Sigma - 1} \Phi(u)
\]
which serve to project onto the appropriate $u^{\pm \im \Sigma}$ basis functions. We use the following integral identities (derivation is presented in the appendix~\ref{app.gamma}):
\begin{align} 
\int_{-\infty}^{+\infty} d u \, \theta(u)\, u^{\im  \Sigma -1} e^{\pm \im \Lambda u} 
&= \int_{0}^{\infty} d u\, u^{\im \Sigma -1} e^{\pm \im \Lambda u}
= \Lambda^{-\im  \Sigma} e^{\mp \frac{\pi \Sigma}{2}} \Gamma(\im \Sigma)\,, \label{KappaRindtoPW-integrals}
\end{align}
where $\Lambda > 0$. Moreover, one has
\begin{align} 
\int_{-\infty}^{+\infty} d u \, \theta(u)\, u^{\im  \Sigma -1} 
&= \int_{0}^{\infty} d u\, u^{\im \Sigma -1}
=2\pi \delta(\Sigma)\,, \label{KappaRindtoPW-integrals-delta}
\end{align}
where it can be verified by using $u=e^x$, and thus $\int_{-\infty}^{+\infty} d x e^{\im x \Sigma}=2\pi \Sigma$.

With the above considerations, one may write (\ref{KR-KPW1}) as
\begin{align} 
&\int_{-\infty}^{+\infty}  d u\,\, \theta(u) u^{\im \Sigma-1} \Phi(u)  \nn\\
&=\int_{-\infty}^{+\infty} d \Omega \frac{2 \pi}{\sqrt{8 \pi \Omega \sinh \left(\frac{\pi \Omega}{\kappa}\right)}}\left(\delta(\Omega+\Sigma) e^{\frac{-\pi \Omega}{2 \kappa}} 
{\cal B}_{\Omega, \kappa}
+\delta(\Omega-\Sigma) e^{-\frac{\pi \Omega}{2 \kappa}} 
{\cal B}_{\Omega, \kappa}^{\dagger}\right) \nn\\
&=\frac{2 \pi}{\sqrt{8 \pi \Sigma \sinh \left(\frac{\pi \Sigma}{\kappa}\right)}}\left(e^{\frac{\pi \Sigma}{2 \kappa}} 
{\cal B}_{-\Sigma, k}+e^{-\frac{\pi \Sigma}{2 \kappa}} 
{\cal B}_{\Sigma, \kappa}^{\dagger}\right)   \\
& =\int_0^{\infty} d \Lambda \frac{1}{\sqrt{8 \pi \Lambda \sinh \left(\frac{\pi \Lambda}{\kappa^{\prime}}\right)}} 
\Bigg\{ \phantom{+} \left( 
e^{-\frac{\pi \Lambda}{2 \kappa^{\prime}}} \Lambda^{-\im \Sigma} e^{-\frac{\pi \Sigma}{2}} \Gamma(\im \Sigma)
+e^{\frac{\pi \Lambda}{2 \kappa^{\prime}}}
\Lambda^{-\im \Sigma} e^{\frac{\pi \Sigma}{2}} \Gamma(\im \Sigma)\right)
{\cal A}_{\Lambda, \kappa'} \nn\\
&\phantom{ =\int_0^{\infty} d \Lambda \frac{1}{\sqrt{8 \pi \Lambda \sinh \left(\frac{\pi \Lambda}{\kappa^{\prime}}\right)}} 
\Bigg\{}
+\left(
e^{-\frac{\pi \Lambda}{2 \kappa^{\prime}}} \Lambda^{-\im \Sigma} e^{\frac{\pi \Sigma}{2}} \Gamma(\im \Sigma)
+e^{\frac{\pi \Lambda}{2 \kappa^{\prime}}}
\Lambda^{-\im \Sigma} e^{\frac{-\pi \Sigma}{2}} \Gamma(\im \Sigma)\right)
{\cal A}_{\Lambda, \kappa'}^{\dagger} \Bigg\} \,, \nn
\end{align}
Next, one may find a linear combination of the above relation and its Hermitian conjugate to find the final result as follows
\setlength\fboxsep{15pt}
\begin{empheq}[box=\fbox]{align}
{\cal B}_{\Omega, \kappa}=\frac{1}{2 \pi} \operatorname{sgn}(\Omega) \int_0^{\infty} &d \Lambda \sqrt{\frac{\Omega}{\sinh \left(\frac{\pi \Omega}{\kappa}\right) \Lambda \sinh \left(\frac{\pi \Lambda}{\kappa^{\prime}}\right)}} \Lambda^{\im \Omega} \Gamma(-\im \Omega)  \label{KappaRindler.intermsof.KappaPlane}\\
&\times\Bigg\{\,\left[
e^{\frac{\pi \Lambda}{2 \kappa^{\prime}}} \sinh \left(\frac{\pi \Omega}{2}(\frac{1}{\kappa}+1)\right)+e^{-\frac{\pi \Lambda}{2 \kappa^{\prime}}} \sinh \left(\frac{\pi \Omega}{2}(\frac{1}{\kappa}-1)\right)\right] {\cal A}_{\Lambda, \kappa'}\nn\\
&\phantom{\times\{} 
+ \left[
e^{\frac{\pi \Lambda}{2 \kappa^{\prime}}} \sinh \left(\frac{\pi \Omega}{2}(\frac{1}{\kappa}-1)\right)+e^{-\frac{\pi \Lambda}{2 \kappa^{\prime}}} \sinh \left(\frac{\pi \Omega}{2}(\frac{1}{\kappa}+1)\right)\right] {\cal A}^{\dagger}_{\Lambda, \kappa'}
\Bigg\},    \nn
\end{empheq}

Two Bogoliubov transformations (\ref{KappaPlane.intermsof.KappaRindler}) and (\ref{KappaRindler.intermsof.KappaPlane}) generalizes the canonical Bogoliubov transformations between Minkowski, Unruh, and Rindler quantizations. It maps the annihilation operator ${\cal A}_{\Lambda,\kappa}$ associated with kappa plane wave modes to the corresponding ${\cal B}_{\Omega,\kappa'}$ operators of kappa Rindler modes in (\ref{KappaPlane.intermsof.KappaRindler}), and, ${\cal B}_{\Omega, \kappa}$  associated with kappa Rindler   modes to the corresponding ${\cal A}_{\Lambda, \kappa'}$ operators of kappa plane wave modes in (\ref{KappaRindler.intermsof.KappaPlane}). Due to its universal structure and capacity to interpolate smoothly between all previously studied mode decompositions, including the Unruh as $\kappa'=1$, Rindler  as $\kappa' \to 0$ and Minkowski plane wave as  $\kappa\to 0$ in (\ref{KappaRindler.intermsof.KappaPlane}), and the Unruh as $\kappa=1$, Rindler  as $\kappa \to 0$, and Minkowski plane wave as  $\kappa' \to 0$  in (\ref{KappaRindler.intermsof.KappaPlane}), we refer to these two equations as the \textit{Mother of All Bogoliubov Transformations}. They  encapsulate the full $\kappa$-deformed quantum map structure and highlights how thermal, observer-dependent vacuum structure arises from underlying analytic squeezing. We summarize all of these special cases in the table~\ref{tab:MOABT} below.

\begin{table}
\centering
\renewcommand{\arraystretch}{1.5}
\begin{tabular}{|c|c|c|l|}
\hline
\textbf{Equation} & \textbf{Limit} & \textbf{Recovered Modes} & \textbf{Operator Mapping} \\
\hline
{(\ref{KappaPlane.intermsof.KappaRindler})} 
& $\kappa' = 1$ & Unruh mode & ${\cal B}_{\Omega,1} = A_\Omega$ for $\Omega \in \mathbb{R}$ \\
& $\kappa' \to 0$ & Rindler mode & ${\cal B}_{\Omega,0} = b_{R\Omega}$, ${\cal B}_{-\Omega,0} = b_{L\Omega}$ for $\Omega > 0$ \\
& $\kappa \to 0$ & Minkowski plane wave & ${\cal A}_{\Lambda, 0} = a_\Lambda$ for $\Lambda > 0$ \\
\hline
{(\ref{KappaRindler.intermsof.KappaPlane})}
& $\kappa = 1$ & Unruh mode & ${\cal B}_{\Omega,1} = A_\Omega$ for $\Omega \in \mathbb{R}$ \\
& $\kappa \to 0$ & Rindler mode & ${\cal B}_{\Omega,0} = b_{R\Omega}$, ${\cal B}_{-\Omega,0} = b_{L\Omega}$  for $\Omega > 0$ \\
& $\kappa' \to 0$ & Minkowski plane wave & ${\cal A}_{\Lambda, 0} = a_\Lambda$  for $\Lambda > 0$ \\
\hline
\end{tabular}
\caption{Limiting behavior of the two Bogoliubov transformations and their operator identifications. These universal relations interpolate between Minkowski, Rindler, and Unruh quantizations, and hence are collectively termed the \textit{Mother of All Bogoliubov Transformations}.}
\label{tab:MOABT}
\end{table}

\section{Commutation relations for different kappa} \label{com.rel.kappa.planewave}

To compute the commutation relations between creation and annihilation operators with different values of $\kappa$, we analyze the inner products of modes defined in (\ref{kappamodeplanewave}). Using the inner product identity (\ref{inner-product}), the overlap between two positive norm modes becomes
\begin{align}
\big\langle \Phi(u,\Lambda,\kappa),\Phi(u,\Lambda',\kappa')\big\rangle =&
 \fft1{{\sqrt{8 \pi \Lambda\, \sinh{(\fft{\pi \Lambda}{\kappa} )}}}}
 \fft1{{\sqrt{8 \pi \Lambda'\, \sinh{(\fft{\pi \Lambda'}{\kappa'} )}}}}  
 \label{innerprod.diff.kappa1} \\
&\times\left\langle e^{\fft{\pi \Lambda}{2\kappa}}\,
e^{-\im \Lambda u}
+e^{-\fft{\pi \Lambda}{2\kappa}}\,e^{\im \Lambda u}\,,\,
e^{\fft{\pi \Lambda'}{2\kappa'}}\,
e^{-\im \Lambda' u}
+e^{-\fft{\pi \Lambda'}{2\kappa'}}\,e^{\im \Lambda' u}
\right\rangle \nn\\
=&\fft{\sinh{\Big(\ft{\pi \Lambda}2\,(\ft1{\kappa}+\ft1{\kappa'} \big)}\Big)}
{\sqrt{{\sinh{(\fft{\pi \Lambda}{\kappa} )}\sinh{(\fft{\pi \Lambda}{\kappa'} )}}}}
\,\delta(\Lambda-\Lambda')\,, \nn
\end{align}
where we have used (\ref{inner-product}). Therefore the commutation relation between annihilation and creation operators with different $\kappa$ and $\kappa'$ is
\begin{align}
\left[{\cal A}_{\Lambda, \kappa},
{\cal A}^\dagger_{\Lambda', \kappa'}\right]=&
\big\langle \Phi(u,\Lambda,\kappa),\Phi(u,\Lambda',\kappa')\big\rangle \nn\\
=&
\fft{
\sinh{\Big(\ft{\pi \Lambda}2\,(\ft1{\kappa}+\ft1{\kappa'} \big)}\Big)}{\sqrt{{\sinh{(\fft{\pi \Lambda}{\kappa} )}\sinh{(\fft{\pi \Lambda}{\kappa'} )}}}}\,\,
\delta(\Lambda-\Lambda') \,.\label{commreldiffkappa1}
\end{align}
As expected, setting $\kappa = \kappa'$ reduces the expression to the standard relation 
\begin{align}
 \left[{\cal A}_{\Lambda, \kappa}, {\cal A}^\dagger_{\Lambda', \kappa}\right]=\delta(\Lambda-\Lambda').   
\end{align}

Next, consider the inner product between a positive norm mode and the complex conjugate of another:
\begin{align}
\big\langle \Phi(u,\Lambda,\kappa) ,\Phi^*(u,\Lambda',\kappa')\big\rangle =&
 \fft1{{\sqrt{8 \pi \Lambda\, \sinh{(\fft{\pi \Lambda}{\kappa} )}}}}
 \fft1{{\sqrt{8 \pi \Lambda'\, \sinh{(\fft{\pi \Lambda'}{\kappa'} )}}}}  \\\label{innerprod.diff.kappa2}
&\times \left\langle e^{\fft{\pi \Lambda}{2\kappa}}\,
e^{-\im \Lambda u}
+e^{-\fft{\pi \Lambda}{2\kappa}}\,e^{\im \Lambda u}\,,\,
e^{\fft{\pi \Lambda'}{2\kappa'}}\,
e^{\im \Lambda' u}
+e^{-\fft{\pi \Lambda'}{2\kappa'}}\,e^{-\im \Lambda' u}
\right\rangle \nn\\
=&\fft{\sinh{\Big(\ft{\pi \Lambda}2\,(\ft1{\kappa}-\ft1{\kappa'} \big)}\Big)}{\sqrt{{\sinh{(\fft{\pi \Lambda}{\kappa} )}\sinh{(\fft{\pi \Lambda}{\kappa'} )}}}}
\,\,
\delta(\Lambda-\Lambda')\,, \nn
\end{align}
where again (\ref{inner-product}) has been used. The commutation relation between annihilation operators with different $\kappa$ and $\kappa'$ is thus
\begin{align}
\left[{\cal A}_{\Lambda, \kappa},
{\cal A}_{\Lambda', \kappa'}\right]=&
-\big\langle \Phi(u,\Lambda,\kappa),\Phi^*(u,\Lambda',\kappa')\big\rangle \nn\\
=&
\fft{
\sinh{\Big(\ft{\pi \Lambda}2\,(\ft1{\kappa'}-\ft1{\kappa} \big)}\Big)}{\sqrt{{\sinh{(\fft{\pi \Lambda}{\kappa} )}\sinh{(\fft{\pi \Lambda}{\kappa'} )}}}}\,\,
\delta(\Lambda-\Lambda') \,.   
\end{align}
Again, setting $\kappa = \kappa'$ gives 
$\left[{\cal A}_{\Lambda, \kappa}, {\cal A}_{\Lambda', \kappa}\right] = 0$, as expected.

These commutation relations can also be derived directly from the Bogoliubov transformation (\ref{bogolfinal}), along with the standard canonical relations:
\begin{align}
 \big[{\cal A}_{\Lambda, \kappa},
{\cal A}^\dagger_{\Lambda', \kappa}\big]=\delta(\Lambda-\Lambda'), \quad \text{and}  \quad  
\big[{\cal A}_{\Lambda, \kappa},
{\cal A}_{\Lambda', \kappa}\big]=0.
\end{align}
as follows.
\begin{align}  
[{\cal A}_{\Lambda,\kappa},{\cal A}_{\Lambda',\kappa'}]
=&\fft{\sinh{\Big(\ft{\pi \Lambda'}2\,(\ft1{\kappa'}-\ft1{\kappa} \big)}\Big)}{\sqrt{{\sinh{(\fft{\pi \Lambda'}{\kappa} )}\sinh{(\fft{\pi \Lambda'}{\kappa'} )}}}} 
\Big[{\cal A}_{\Lambda,\kappa}, {\cal A}^{\dagger}_{\Lambda',\kappa} 
\Big] \nn\\
=&
\fft{\sinh{\Big(\ft{\pi \Lambda}2\,(\ft1{\kappa'}-\ft1{\kappa} \big)}\Big)}{\sqrt{{\sinh{(\fft{\pi \Lambda}{\kappa} )}\sinh{(\fft{\pi \Lambda}{\kappa'} )}}}}
\delta(\Lambda-\Lambda')
\,, \nn\\
[{\cal A}_{\Lambda,\kappa},{\cal A}^{\dagger}_{\Lambda',\kappa'}]
=&\fft{\sinh{\Big(\ft{\pi \Lambda'}2\,(\ft1{\kappa'}+\ft1{\kappa} \big)}\Big)}{\sqrt{{\sinh{(\fft{\pi \Lambda'}{\kappa} )}\sinh{(\fft{\pi \Lambda'}{\kappa'} )}}}} 
\Big[{\cal A}_{\Lambda,\kappa}, {\cal A}^{\dagger}_{\Lambda',\kappa} 
\Big]\nn\\
=&
\fft{\sinh{\Big(\ft{\pi \Lambda}2\,(\ft1{\kappa'}+\ft1{\kappa} \big)}\Big)}{\sqrt{{\sinh{(\fft{\pi \Lambda}{\kappa} )}\sinh{(\fft{\pi \Lambda}{\kappa'} )}}}}
\delta(\Lambda-\Lambda')
\,, \label{commrel}
\end{align}
\section{Continuous squeezing} \label{sec.cont.squeezing}

\subsection{Relation between distinct kappa plane wave vacua}

Distinct $\kappa$-vacua can be related to each other through a transformation structurally similar to the Minkowski–Rindler relation. To obtain the relation between two $\kappa$-vacua, consider that
\begin{equation}
{\cal A}_{\Lambda,\kappa'} \ket{0_{\kappa'}} = 0.
\end{equation}
Then, using the Bogoliubov transformation in (\ref{bogolfinal}), it follows that
\begin{equation}
\left({\cal A}_{\Lambda,\kappa} - \eta_{\kappa,\kappa',\Lambda}\, {\cal A}^{\dagger}_{\Lambda,\kappa} \right)
\ket{0_{\kappa'}} = 0\,, \label{Kappa.PW.vs.Kappa.PW.vac1}
\end{equation}
where the squeezing parameter $\eta_{\kappa,\kappa',\Lambda}$ is defined as
\begin{equation}
\eta_{\kappa,\kappa',\Lambda} = \frac{\sinh{\left(\frac{\pi \Lambda}{2}\left(\frac{1}{\kappa} - \frac{1}{\kappa'}\right)\right)}}
{\sinh{\left(\frac{\pi \Lambda}{2}\left(\frac{1}{\kappa} + \frac{1}{\kappa'}\right)\right)}}\,. \label{eta}
\end{equation}
Note that the parameter $\eta_{\kappa,\kappa',\Lambda}$ satisfies the bound $|\eta_{\kappa,\kappa',\Lambda}| < 1$, which is required for the squeezing relation $\left(a - \alpha a^\dagger \right) \ket{\psi_\alpha} = 0$ to uniquely characterize a squeezed vacuum, as discussed in \cite{Azizi2025Uniqueness}. This inequality is ensured because for $\Lambda > 0$, $\kappa > 0$, and $\kappa' > 0$, we always have
\begin{equation}
\frac{\pi \Lambda}{2}\left(\frac{1}{\kappa} + \frac{1}{\kappa'}\right) 
> \left| \frac{\pi \Lambda}{2}\left(\frac{1}{\kappa} - \frac{1}{\kappa'}\right) \right|.
\end{equation}
Since $\sinh(x)$ is strictly increasing for $x > 0$ and satisfies $|\sinh(-x)| = \sinh(x)$, it follows that the denominator of (\ref{eta}) is always greater than the numerator, hence
\begin{equation}
|\eta_{\kappa,\kappa',\Lambda}| < 1,
\end{equation}
confirming the physical validity of the continuous squeezing structure connecting different $\kappa$-vacua.

Using our result in Appendix \ref{app.2forms}, we find
\setlength\fboxsep{15pt}
\begin{empheq}[box=\fbox]{align}
\ket{0_{\kappa'}}=&\fft1{\sqrt{Z_{\kappa\kappa'}}}\,\,
\exp{\int_{0}^{\infty} d\Lambda\, \ft12\eta_{\kappa,\kappa',\Lambda}
\,{\cal A}^{\dagger}_{\Lambda, \kappa}\,{\cal A}^{\dagger}_{\Lambda, \kappa}}\, \ket{0_{\kappa}}\nn\\
=&
\exp{\int_{0}^{\infty} d\Lambda\,\ft12  \tanh^{-1}{\eta_{\kappa,\kappa',\Lambda}}\,
\Big(
{\cal A}_{\Lambda, \kappa}\,{\cal A}_{\Lambda, \kappa}
-{\cal A}^{\dagger}_{\Lambda, \kappa}\,{\cal A}^{\dagger}_{\Lambda, \kappa}
\Big)}
\, \ket{0_{\kappa}},
\label{Kappa.PW.vs.Kappa.PW.vac2}
\end{empheq}
where $Z_{\kappa\kappa'}$ is the normalization factor depending on $\kappa$ and $\kappa'$, as defined in (\ref{twoforms}). Note that the second line of (\ref{Kappa.PW.vs.Kappa.PW.vac2}) is expressed as the action of a unitary squeezing operator on $\ket{0_\kappa}$, and hence the resulting state $\ket{0_{\kappa'}}$ is automatically normalized.

\subsection{Kappa plane vacuum in terms of Minkowski}
Let us now express the general $\kappa$-plane wave vacuum in terms of the Minkowski vacuum. This can be readily obtained by taking the limit $\kappa \to 0^+$ in (\ref{Kappa.PW.vs.Kappa.PW.vac2}). 

First, observe from (\ref{eta}) that $\eta_{\kappa,\kappa',\Lambda}$ is an odd function under the exchange $(\kappa \leftrightarrow \kappa')$. To evaluate the limit $\kappa \to 0^+$, we use the asymptotic behavior $\sinh(z) \sim \pm e^{\pm z}/2$ for large $|z|$. In this regime, the ratio becomes
$$\eta_{\kappa,\kappa',\Lambda}= \frac{\sinh{\left(\frac{\pi \Lambda}{2\kappa} - \frac{\pi \Lambda}{2\kappa'}\right)}}{\sinh{\left(\frac{\pi \Lambda}{2\kappa} + \frac{\pi \Lambda}{2\kappa'}\right)}} \approx \frac{e^{\frac{\pi \Lambda}{2\kappa} - \frac{\pi \Lambda}{2\kappa'}}}{e^{\frac{\pi \Lambda}{2\kappa} + \frac{\pi \Lambda}{2\kappa'}}}=e^{-\frac{\pi \Lambda}{\kappa'}}. $$
Therefore, we obtain the limiting expression
\begin{equation}
\eta_{0^+,\kappa',\Lambda}=
-\eta_{\kappa',0^+,\Lambda}
\to e^{-\frac{\pi \Lambda}{\kappa'}} \label{eta.1}
\end{equation}
Since ${\cal A}_{\Lambda,0^+} = a_{\Lambda}$, and by relabeling $\kappa' \to \kappa$, and $\Lambda \to \nu$, equation (\ref{Kappa.PW.vs.Kappa.PW.vac1}) reduces to
\begin{equation}
\left(a_{\nu} - e^{-\frac{\pi \nu}{\kappa}} a^{\dagger}_{\nu} \right)
\ket{0_{\kappa}} = 0\,, \label{Kappa.PW.vs.Mink.Bog}
\end{equation}
and (\ref{Kappa.PW.vs.Kappa.PW.vac2}) yields
\setlength\fboxsep{15pt}
\begin{empheq}[box=\fbox]{align}
\ket{0_{\kappa}}=&\fft1{\sqrt{Z_{\kappa}}}\,\,
\exp{\ft12\int_{0}^{\infty} d\nu\,
e^{-\frac{\pi \nu}{\kappa}}\,
a^{\dagger}_{\nu}\,a^{\dagger}_{\nu}}\, \ket{0_M}\nn\\
=& 
\exp{-\ft14\int_{0}^{\infty} d\nu\,\ln\left( \coth\left( \frac{\pi \nu}{2\kappa} \right) \right)\,
\Big(
a^{\dagger}_{\nu}\,a^{\dagger}_{\nu}
-a_{\nu}\,a_{\nu}
\Big)}\ket{0_M},
\label{Kappa.PW.vs.Mink.vac}
\end{empheq}
where ${Z}_{\kappa}$ is a normalization factor. 

To gain more intuition about this result, define $\gamma = e^{-\frac{\pi \nu}{\kappa}}$ and focus on a single-frequency mode $\nu$. The corresponding vacuum state becomes
\begin{equation}
\ket{0_{\kappa}} = \frac{1}{\sqrt{Z_{\kappa}}}
\sum_{n=0}^{\infty} \frac{1}{n!} \left( \frac{\gamma}{2} \right)^n 
\left(a^{\dagger}_{\nu}\right)^{2n} \ket{0_M}\,.
\end{equation}
Using the identity $(a^{\dagger}_{\nu})^n \ket{0_M} = \sqrt{n!} \ket{n}_{\nu}$, we arrive at
\begin{equation}
\ket{0_{\kappa}} = \frac{1}{\sqrt{Z_{\kappa}}}
\sum_{n=0}^{\infty} \sqrt{\binom{2n}{n}} \left(\frac{\gamma}{2} \right)^n 
\ket{2n}_{\nu}\,, \label{kappavac-expanded}
\end{equation}
demonstrating that the $\kappa$-vacuum is a squeezed vacuum composed exclusively of even-number Fock states with respect to Minkowski plane wave modes.

\subsection{Interpretation as a continuous-mode squeezed vacuum}

The derivation above reveals that the $\kappa$-plane wave vacuum $\ket{0_\kappa}$ is annihilated by a Bogoliubov-transformed operator for each frequency $\nu > 0$:
\begin{align}
\left( a_\nu - e^{-\frac{\pi \nu}{\kappa}} a^\dagger_\nu \right) \ket{0_\kappa} = 0,
\end{align}
where $a_\nu$ and $a^\dagger_\nu$ denote standard Minkowski plane wave operators. This is the hallmark of a squeezed vacuum state, with a frequency-dependent squeezing parameter $r(\nu)$ associated to each mode. Kappa plane wave vacuum, as stated in (\ref{Kappa.PW.vs.Mink.vac}) can be written as:
\begin{align}
\boxed{\quad \ket{0_\kappa} 
= \exp\left\{ -\frac{1}{2} \int_0^\infty d\nu\, r(\nu) 
\left( a^\dagger_\nu a^\dagger_\nu - a_\nu a_\nu \right) \right\} \ket{0}, \,\,\, \text{with} \,\, \, r(\nu) = \frac{1}{2} \ln \coth\left( \frac{\pi \nu}{2\kappa} \right), \quad }
\end{align}
which is precisely the form of a continuous-mode squeezed vacuum state.  

Continuous squeezing is a well-established concept in quantum optics and continuous-variable quantum information theory, where multimode bosonic fields are manipulated to produce squeezed states with frequency-dependent squeezing parameters \cite{Braunstein2005, Weedbrook2012}. Such states, described by Gaussian operations on a continuum of modes, are pivotal in applications ranging from quantum metrology to entanglement generation \cite{ScullyZubairy1997}. In this work, we demonstrate that the $\kappa$-plane wave vacuum $\ket{0_\kappa}$ is a continuous-mode squeezed vacuum, characterized by a squeezing parameter $r(\nu) = \frac{1}{2} \ln \coth\left( \frac{\pi \nu}{2\kappa} \right)$. Unlike traditional applications in optical systems, our construction applies continuous squeezing to quantum field theory in flat spacetime, defining a tunable family of vacua that interpolate to the Minkowski vacuum as $\kappa \to 0$. This novel application bridges Gaussian quantum information with relativistic quantum field theory, offering new insights into vacuum structure and observer-dependent effects.
This interpretation places $\ket{0_\kappa}$ within the broader family of Gaussian states familiar in quantum optics and continuous-variable quantum information theory \cite{Braunstein2005, Weedbrook2012}. The deformation parameter $\kappa$ shapes the squeezing spectrum: low-frequency modes exhibit strong squeezing, while high-frequency modes asymptotically resemble those of the Minkowski vacuum. This formalism opens the door to leveraging Gaussian techniques for analyzing entanglement, thermality, and correlations in $\kappa$-vacua.

\subsection{Minkowski vacuum in terms of kappa plane vacuum}

Let us now express the Minkowski vacuum in terms of the $\kappa$-plane wave vacuum. This can be directly obtained by taking the limit $\kappa' \to 0$ in (\ref{Kappa.PW.vs.Kappa.PW.vac2}). Using the asymptotic behavior previously discussed, we find
\begin{equation}
\eta_{\kappa,0^+,\Lambda} \to -e^{-\frac{\pi \Lambda}{\kappa}}\,. \label{eta.2}
\end{equation}

Substituting this into the squeezed vacuum condition (\ref{Kappa.PW.vs.Kappa.PW.vac1}) gives the Bogoliubov relation
\begin{equation}
\left({\cal A}_{\Lambda,\kappa} +
e^{-\frac{\pi \Lambda}{\kappa}}\, {\cal A}^{\dagger}_{\Lambda,\kappa} \right)
\ket{0_M} = 0\,, \label{Mink.vs.Kappa.PW.Bog}
\end{equation}
which characterizes the Minkowski vacuum as a squeezed state over $\ket{0_\kappa}$.

Accordingly, the transformation (\ref{Kappa.PW.vs.Kappa.PW.vac2}) becomes
\setlength\fboxsep{15pt}
\begin{empheq}[box=\fbox]{align}
\ket{0_M}
=&\frac{1}{\sqrt{\widetilde{Z}_{\kappa}}}
\exp\left( -\frac{1}{2} \int_{0}^{\infty} d\Lambda\,
e^{-\frac{\pi \Lambda}{\kappa}}\, 
{\cal A}^{\dagger}_{\Lambda, \kappa}\,{\cal A}^{\dagger}_{\Lambda, \kappa} \right) \ket{0_{\kappa}} \nn\\
=&
\exp\left( \frac{1}{4} \int_{0}^{\infty} d\Lambda\, \ln\left( \coth\left( \frac{\pi \Lambda}{2\kappa} \right) \right)
\left(
{\cal A}_{\Lambda, \kappa}\,{\cal A}_{\Lambda, \kappa}
-{\cal A}^{\dagger}_{\Lambda, \kappa}\,{\cal A}^{\dagger}_{\Lambda, \kappa}
\right) \right) \ket{0_{\kappa}},
\label{Mink.vs.Kappa.PW.vac}
\end{empheq}
where $\widetilde{Z}_{\kappa}$ is a normalization constant. This confirms that the Minkowski vacuum can be constructed from the $\kappa$-vacuum by a continuous multimode squeezing operation.

\section{Conclusion} \label{conclusion}

In this work, we introduced a new class of mode functions in flat spacetime—called $\kappa$-plane wave modes—formed by $\kappa$-dependent linear combinations of positive and negative frequency Minkowski plane waves. These modes define a one-parameter family of vacua $\ket{0_\kappa}$ that reduce to the Minkowski vacuum as $\kappa \to 0$, but do not recover the Rindler vacuum in any limit. This contrasts with earlier $\kappa$-Rindler constructions and highlights the $\kappa$-plane wave quantization as a distinct vacuum structure.

A central finding of this paper is that $\ket{0_\kappa}$ is uniquely characterized as a continuous-mode squeezed vacuum. We showed that the condition
\[
\left( a_\nu - e^{-\frac{\pi \nu}{\kappa}} a^\dagger_\nu \right) \ket{0_\kappa} = 0
\]
yields a squeezing parameter $r(\nu)$ via $\tanh r(\nu) = e^{-\frac{\pi \nu}{\kappa}}$, identifying $\ket{0_\kappa}$ as a pure Gaussian state with mode-dependent squeezing. This construction generalizes naturally to the continuous-frequency limit and connects the $\kappa$-vacua to well-studied structures in continuous-variable quantum field theory.

We also found that operators with different $\kappa$ values obey nontrivial commutation relations, revealing that $\kappa$ plays the role of a continuous deformation parameter that governs vacuum entanglement and mode structure.

These insights provide a new lens for viewing vacuum structure in quantum field theory. The $\kappa$-vacua describe a continuum of squeezed Gaussian states that interpolate between pure and thermal-like behavior. This opens up a range of possibilities for applications in analog gravity, detector response theory, and experimental simulations of non-inertial field dynamics.

\section*{Acknowledgments}

I am grateful to Girish Agarwal, Reed Nessler, Marlan Scully, Philip Stamp, Anatoly Svidzinsky, and Bill Unruh for illuminating discussions. I would also like to extend special thanks to an anonymous referee for an insightful question  concerning the uniqueness of the mode coefficients. This work was supported by the Robert A. Welch Foundation (Grant No. A-1261) and the National Science Foundation (Grant No. PHY-2013771).

\appendix 
\section{Notations} \label{notation}

Throughout this paper, we use the notation $\Lambda \in \mathbb{R}^+$ to denote the frequency label for the $\kappa$-plane wave modes. The operator ${\cal A}_{\Lambda, \kappa}$ represents the annihilation operator associated with a $\kappa$-plane wave mode. On the other hand, $\Omega \in \mathbb{R}$ denotes the frequency parameter used for $\kappa$-Rindler modes, following the convention introduced in \cite{Azizi2023}. The corresponding annihilation operators for the $\kappa$-Rindler modes are denoted by ${\cal B}_{\Omega, \kappa}$.

\section{On the Uniqueness and Motivation of the Kappa-Plane Wave Ansatz}
\label{app.ansatz}

In this appendix, we provide a more detailed justification for the specific form of the \(\kappa\)-plane wave ansatz. We first address the phases of the mode coefficients and then show that while a general family of solutions to the normalization constraint exists, the chosen ansatz is uniquely required to define a consistent, frequency-independent \(\kappa\) parameter and to yield the desired physical interpretation of continuous squeezing.

Before proceeding, we consider the possibility of the mode coefficients \(\alpha(\Omega)\) and \(\beta(\Omega)\) being complex. While an overall phase can be absorbed by a simple redefinition of the field mode, the \textit{relative phase} between \(\alpha(\Omega)\) and \(\beta(\Omega)\) cannot be eliminated and remains physically significant. For this work, we ignore the phase and consider  \(\alpha(\Omega)\) and \(\beta(\Omega)\) as real and positive numbers.
\subsection*{General Parametrization vs. Specific Ansatz}
With the coefficients taken as real and positive, they must satisfy the normalization constraint:
\begin{equation}
    \alpha(\Omega)^2 - \beta(\Omega)^2 = \frac{1}{4\pi\Omega}.
    \label{eq:norm_constraint_app}
\end{equation}
A general solution can be constructed by parameterizing the coefficients using an arbitrary function \(\lambda(\Omega)\):
\begin{align}
    \alpha(\Omega) = \frac{1}{\sqrt{4\pi\Omega}} \cosh \lambda(\Omega), \qquad
    \beta(\Omega) = \frac{1}{\sqrt{4\pi\Omega}} \sinh \lambda(\Omega).\label{eq:gen_alpha_beta}
\end{align}
This general form satisfies the constraint for any choice of \(\lambda(\Omega)\).

The specific ansatz introduced in the main text Eq.~(\ref{ansatz}) is:
\begin{align}
    \alpha(\kappa, \Omega) = \frac{e^{\frac{\pi\Omega}{2\kappa}}}{\sqrt{8\pi\Omega\sinh(\frac{\pi\Omega}{\kappa})}}, \qquad
    \beta(\kappa, \Omega) = \frac{e^{-\frac{\pi\Omega}{2\kappa}}}{\sqrt{8\pi\Omega\sinh(\frac{\pi\Omega}{\kappa})}}. \label{eq:spec_beta}
\end{align}
Comparing (\ref{eq:gen_alpha_beta}) and (\ref{eq:spec_beta}), we have
\begin{equation}
    \tanh \lambda(\Omega) = e^{-\frac{\pi\Omega}{\kappa}}.
    \label{eq:lambda_choice}
\end{equation}
The parameter \(\kappa\) is a global constant characterizing the vacuum and must be independent of the mode frequency \(\Omega\). Starting from the general ansatz in Eq.~\eqref{eq:gen_alpha_beta} with an arbitrary function \(\lambda(\Omega)\), and using Eq.~\eqref{eq:lambda_choice} to define \(\kappa\), we obtain:
\begin{equation}
\lambda(\Omega) = \tanh^{-1} \left( e^{-\frac{\pi \Omega}{\kappa}} \right) = \frac{1}{2} \ln \coth \left( \frac{\pi \Omega}{2 \kappa} \right), \qquad \kappa(\Omega) = -\frac{\pi \Omega}{\ln \left( \tanh \lambda(\Omega) \right)}.
\end{equation}
For a generic \(\lambda(\Omega)\), the resulting \(\kappa\) depends on \(\Omega\), which contradicts the requirement that \(\kappa\) is a constant deformation parameter. Thus, the ansatz in Eq.~\eqref{ansatz} constrains the choice of \(\lambda(\Omega)\).

\subsection*{Physical Motivation: Continuous Squeezing}

The choice of ansatz is physically motivated. As shown in the main text, this ansatz characterizes the \(\kappa\)-vacuum as a continuous-mode squeezed state of the Minkowski vacuum, a result that directly connects to Gaussian quantum information. The ansatz’s exponential coefficients, \( e^{\pm \frac{\pi \Lambda}{2 \kappa}} \), yield a Bogoliubov transformation linking the \(\mathcal{A}_{\Lambda,\kappa}\) operators to the Minkowski operators \( a_\Lambda \), enabling the squeezed state condition \((a_{\nu} - e^{-\frac{\pi \nu}{\kappa}} a_{\nu}^{\dagger}) |0_{\kappa}\rangle = 0\).

This simple exponential form, inherited from the ansatz, defines the vacuum as a squeezed state with \(\kappa\) as the tunable squeezing parameter. A different coefficient choice would complicate the transformation, obscuring the elegant squeezed state structure and the physical significance of \(\kappa\). Thus, the ansatz is essential for revealing the continuous squeezing structure of the \(\kappa\)-vacuum.

\section{Gamma Integrals} \label{app.gamma}

To evaluate integrals of the form
\begin{align}
\int_{-\infty}^{+\infty} \mathrm{d} u \, \theta(u)\, u^{i \Sigma -1} e^{\pm i \Lambda u},
\end{align}
we use the standard Gamma function identity
\begin{align}
\int_0^\infty u^{s-1} e^{-bu} \, \mathrm{d}u = b^{-s} \Gamma(s),
\end{align}
which holds for $b, s \in \mathbb{C}$ with $\Re(b) > 0$. In our case, we identify $s = i\Sigma$ and $b = \pm i\Lambda + \epsilon$, where $0 < \epsilon \ll 1$ is a small positive regulator ensuring convergence.

To compute $b^{-s}$, we evaluate:
\begin{align}
(i\Lambda)^{-i\Sigma} 
&= \exp\left(-i \Sigma \ln(i \Lambda)\right)
= \exp\left(-i \Sigma \left[\ln \Lambda + \frac{i\pi}{2}\right] \right)
= \Lambda^{-i\Sigma} e^{\frac{\pi \Sigma}{2}}, \nn\\
(-i\Lambda)^{-i\Sigma}
&= \exp\left(-i \Sigma \ln(-i \Lambda)\right)
= \exp\left(-i \Sigma \left[\ln \Lambda - \frac{i\pi}{2} \right] \right)
= \Lambda^{-i\Sigma} e^{-\frac{\pi \Sigma}{2}}.
\end{align}

Therefore, the regulated integrals evaluate to
\begin{align}
\int_{-\infty}^{+\infty} \mathrm{d} u \, \theta(u) u^{i\Sigma - 1} e^{+i\Lambda u} 
&= \int_{0}^{\infty} u^{i\Sigma - 1} e^{+i\Lambda u} \, \mathrm{d}u 
= \Lambda^{-i\Sigma} e^{-\frac{\pi \Sigma}{2}} \Gamma(i\Sigma), \nn\\
\int_{-\infty}^{+\infty} \mathrm{d} u \, \theta(u) u^{i\Sigma - 1} e^{-i\Lambda u}
&= \int_{0}^{\infty} u^{i\Sigma - 1} e^{-i\Lambda u} \, \mathrm{d}u 
= \Lambda^{-i\Sigma} e^{+\frac{\pi \Sigma}{2}} \Gamma(i\Sigma),
\end{align}
where we assume $\Lambda > 0$. Identical results hold for the integrals over $\theta(-u)$.

\section{Two forms: Non-unitary and unitary actions on vacuum} \label{app.2forms}

\subsection{Single frequency}

The goal is to revisit the $\kappa$-plane wave vacuum $\ket{0_\kappa}$ and elucidate its structure, especially in connection with the uniqueness of squeezed vacuum states. We begin with the defining relation:
\begin{align}
\left(a_\nu - e^{-\frac{\pi \nu}{\kappa}} a^\dagger_\nu \right) \ket{0_\kappa} = 0,
\end{align}
where $a_\nu$ and $a^\dagger_\nu$ are Minkowski plane wave operators. The general solution to such constraints is given in \cite{Azizi2025Uniqueness} as
\setlength\fboxsep{15pt}
\begin{empheq}[box=\fbox]{align}
\left(a - \alpha a^\dagger \right) \ket{\psi_\alpha} &= 0
\quad \Rightarrow \quad 
\ket{\psi_\alpha} = \frac{1}{\sqrt{Z}}\, e^{\frac{1}{2} \alpha a^{\dagger 2}} \ket{0}
= \exp\left\{ \frac{1}{2} \left( \xi a^{\dagger 2} - \xi^* a^2 \right) \right\} \ket{0}, \nn\\
& \text{with} \quad \alpha = \tanh r\, e^{i(\theta + \pi)}, \quad \xi = r e^{i\theta}.
\end{empheq}
For real $\alpha$, we have $\theta \in \{0, \pi\}$, which implies $\xi = -\tanh^{-1} \alpha$. Hence,
\begin{align}
\left(a - \alpha a^\dagger \right) \ket{\psi_\alpha} = 0
\quad \Rightarrow \quad 
\ket{\psi_\alpha} = \frac{1}{\sqrt{Z}}\, 
e^{\frac{1}{2} \alpha a^{\dagger 2}} \ket{0}
= \exp\left\{ -\frac{1}{2} \tanh^{-1} \alpha 
\left( a^{\dagger 2} - a^2 \right) \right\} \ket{0}. \label{twoforms}
\end{align}
For $\alpha = e^{-\frac{\pi \Lambda}{\kappa}}$, we obtain
\begin{align}
\tanh^{-1}\left(e^{-\frac{\pi \Lambda}{\kappa}}\right) 
= \frac{1}{2} \ln\left( \coth\left( \frac{\pi \Lambda}{2\kappa} \right) \right)
\quad \Rightarrow \quad 
\xi = -\frac{1}{2} \ln\left( \coth\left( \frac{\pi \Lambda}{2\kappa} \right) \right).
\end{align}

\subsection{Finite frequencies}

A key observation is that for the unique $\kappa$-plane wave vacuum $\ket{0_\kappa}$, the constraint
\begin{align}
\left(a_\nu - e^{-\frac{\pi \nu}{\kappa}} a^\dagger_\nu \right) \ket{0_\kappa} = 0, \label{Cont.squeezing.1}
\end{align}
holds for all $\nu \in \mathbb{R}^+$. We now generalize the construction to multiple or continuous frequency modes.

First, consider a discrete set of $N$ modes with frequencies $\nu_1, \nu_2, \ldots, \nu_N$. The commutation relations are
\begin{align}
[a_{\nu_i}, a^\dagger_{\nu_j}] = \delta_{ij},
\end{align}
and we impose the constraint
\begin{align}
\left(a_{\nu_i} - e^{-\frac{\pi \nu_i}{\kappa}} a^\dagger_{\nu_i} \right) \ket{0_\kappa} = 0 \quad \forall i. \label{Cont.squeezing.2}
\end{align}
This is satisfied by the squeezed state ansatz
\begin{align}
\ket{0_\kappa} 
&= \exp\left\{ \sum_{i=1}^N \frac{1}{2} e^{-\frac{\pi \nu_i}{\kappa}} a^{\dagger 2}_{\nu_i} \right\} \ket{0} \\
&= \exp\left\{ \sum_{i=1}^N \frac{-r_i}{2} \left( a^{\dagger 2}_{\nu_i} - a^2_{\nu_i} \right) \right\} \ket{0}, \label{**}
\end{align}
where $r_i = \frac{1}{2} \ln \coth \left( \frac{\pi \nu_i}{2\kappa} \right)$.
Defining the squeezing operator as
\begin{align}
S = \prod_{i=1}^N S_i, \quad 
S_i = \exp\left( \frac{-r_i}{2} (a^{\dagger 2}_{\nu_i} - a^2_{\nu_i}) \right),
\end{align}
we verify that each $[a_{\nu_i}, S_j] = 0$ for $i \ne j$, and
\begin{align}
[a_{\nu_i}, S] = S_1 \cdots [a_{\nu_i}, S_i] \cdots S_N.
\end{align}
Hence, $\ket{0_\kappa} = S \ket{0}$ satisfies the squeezing condition.

\subsection{Continuous frequencies}

Extending to the continuum limit, we take
\begin{align}
[a_{\nu}, a^\dagger_{\nu'}] = \delta(\nu - \nu'),
\end{align}
and postulate the general form
\begin{align}
\ket{0_\kappa} = \frac{1}{\sqrt{Z_\kappa}} \,
\exp\left( \int_0^\infty d\nu' d\nu''\, \alpha(\nu', \nu'')\, a^\dagger_{\nu'} a^\dagger_{\nu''} \right) \ket{0_M}, \label{Kappa planewave-Mink.vacua2}
\end{align}
where $\alpha(\nu,\nu')$ is symmetric.

Computing the commutator yields
\begin{align}
&\left[a_\nu, 
\exp\left( \int_0^{\infty} d\nu' d\nu''\, \alpha(\nu',\nu'')\, a^{\dagger}_{\nu'} a^{\dagger}_{\nu''} \right) \right] \nn\\
&= \int_0^{\infty} d\nu' d\nu''\, \alpha(\nu',\nu'') 
\left( a^{\dagger}_{\nu'}\, \delta(\nu - \nu'') + a^{\dagger}_{\nu''}\, \delta(\nu - \nu') \right)
\exp\left( \int_0^{\infty} d\nu' d\nu''\, \alpha(\nu',\nu'')\, a^{\dagger}_{\nu'} a^{\dagger}_{\nu''} \right) \nn\\
&= \int_0^{\infty} d\nu'\, 2\, \alpha(\nu,\nu')\, a^{\dagger}_{\nu'}\,
\exp\left( \int_0^{\infty} d\nu' d\nu''\, \alpha(\nu',\nu'')\, a^{\dagger}_{\nu'} a^{\dagger}_{\nu''} \right)\,. \label{Kappa planewave-Mink.vacua4}
\end{align}

Choosing 
\begin{align}
\alpha(\nu,\nu') = \frac{1}{2} e^{-\frac{\pi \nu}{\kappa}} \delta(\nu - \nu'),
\end{align}
we obtain

\begin{equation}
\left[a_\nu, 
\exp\left( \ft12 \int_0^{\infty} d\nu'\,  e^{-\frac{\pi \nu'}{\kappa}} \left(a^{\dagger}_{\nu'}\right)^2 \right) \right]
= e^{-\frac{\pi \nu}{\kappa}}\, a^{\dagger}_{\nu}
\exp\left( \ft12 \int_0^{\infty} d\nu'\,  e^{-\frac{\pi \nu'}{\kappa}} \left(a^{\dagger}_{\nu'}\right)^2 \right)\,.
\end{equation}
Hence, applying $a_\nu$ to the exponential acting on the Minkowski vacuum yields
\begin{align}
a_\nu  
\exp\left( \ft12 \int_0^{\infty} d\nu'\,  e^{-\frac{\pi \nu'}{\kappa}} \left(a^{\dagger}_{\nu'}\right)^2 \right) \ket{0_M}
&= \left[a_\nu, 
\exp\left( \ft12 \int_0^{\infty} d\nu'\,  e^{-\frac{\pi \nu'}{\kappa}} \left(a^{\dagger}_{\nu'}\right)^2 \right) \right] \ket{0_M} \nn\\
&= e^{-\frac{\pi \nu}{\kappa}}\, a^{\dagger}_{\nu}
\exp\left( \ft12 \int_0^{\infty} d\nu'\,  e^{-\frac{\pi \nu'}{\kappa}} \left(a^{\dagger}_{\nu'}\right)^2 \right) \ket{0_M}\,. \label{Kappa planewave-Mink.vacua3}
\end{align}
Comparing equation (\ref{Kappa planewave-Mink.vacua3}) with the constraint (\ref{Cont.squeezing.1}), we find that the $\kappa$-plane wave vacuum is given by
\begin{equation}
\ket{0_{\kappa}} = \frac{1}{\sqrt{Z_{\kappa}}}
\exp\left( \ft12 \int_0^{\infty} d\nu\, e^{-\frac{\pi \nu}{\kappa}} \left(a^{\dagger}_{\nu}\right)^2 \right) \ket{0_M}\,, \label{Kappa-vac-final}
\end{equation}
where $Z_{\kappa}$ is the normalization constant.

\bibliographystyle{jhep}
\bibliography{UnruhRef}
\end{document}